\documentclass[showpacs,aps,twocolumn,prl,floatfix]{revtex4}

\usepackage{amssymb}
\usepackage{amsmath}
\usepackage[dvips]{graphicx}
\usepackage{bm}

\usepackage{times}

\newcommand{\smeq}{\! = \!}

\newcommand{\smpl}{\! + \!}
\newcommand{\smmi}{\! - \!}

\newcommand{\Ef}{E_{\text{F}}}

\newcommand{\be}{\begin{equation}}
\newcommand{\ee}{\end{equation}}
\newcommand{\bea}{\begin{eqnarray}}
\newcommand{\eea}{\end{eqnarray}}

\newcommand{\up}{\uparrow}

\newcommand{\ci}{\mathrm{i}}

\begin{document}

\title{Edge states interferometry and spin rotations in zigzag graphene nanoribbons}
\author{Gonzalo Usaj}
\affiliation{Centro At{\'o}mico Bariloche and Instituto Balseiro,
Comisi\'on Nacional de Energ\'{\i}a At\'omica, 8400 S. C. de Bariloche, Argentina}
 \affiliation{Consejo Nacional de
Investigaciones Cient\'{\i}ficas y T\'ecnicas (CONICET), Argentina}
\date{April 29, 2009}

\begin{abstract}
An interesting property of zigzag graphene nanoribbons is the presence of edge states which are extended along its borders but localized in the transverse direction. We show that because of this property, electron transport through an externally induced potential well displays two-paths-interference oscillations when subject either to a magnetic or a transverse electric field. This effect does not require the existence of an actual `hole' in the nanoribbon's geometry. Moreover, since edge states are spin polarized, having opposite polarization on opposite sides, such interference effect can be used to rotate the spin of the incident carriers in a controlled way.

\end{abstract}
\maketitle
Graphene, a two dimensional array of carbon atoms in a honeycomb lattice, is a very interesting material with unusual electronic properties \cite{Katsnelson2006,Novoselov2007,Geim2007,CastroNeto-review}. It has attracted much of attention since its first experimental realization \cite{Novoselov2004,Novoselov2005} as it offers a great  potential for technological applications while, at the same time,  it has lead to the observation of new physical phenomena such as an anomalous quantization of the Hall effect \cite{Zhang2005,Novoselov2007a}, observable at room temperature, or the manifestation of the Klein tunneling paradox in transport \cite{Katsnelson2006,Cheianov2007,Young2009}, among others. The key for understanding graphene's peculiarities relies on its band structure: electronic excitations around the Fermi energy ($\Ef$)  can be described by an effective Hamiltonian that mimics the Dirac equation for massless chiral fermions where the spin is replaced by a pseudospin (the two inequivalent sites of the honeycomb lattice) and the speed of light by the Fermi velocity \cite{CastroNeto-review,Beenakker2008}. The actual spin plays no crucial role in bulk samples.

A novel effect unique to graphene appears in graphene nanoribbons (GNRs): when the termination of the GNR corresponds to a  zigzag ordering of the carbon atoms (see Fig. 1a) the system presents edge states \cite{Fujita1996,Nakada1996,Brey2006}. That is, there are eigenfunctions that are extended along the zigzag nanoribbon (ZGNR), but that decay exponentially away from the edges towards the center of the ZGNR. 
These estates have recently been observed in graphite surfaces near monoatomic step edges \cite{Niimi2006}.
From the theoretical point of view, they can be easily obtained from either a discrete tight-binding model for the honeycomb lattice \cite{Fujita1996,Nakada1996} or a low energy effective Hamiltonian (Dirac equation) \cite{Brey2006}.  
If only nearest neighbors hopping is considered in the former, the edge states have an exponentially small group velocity $v_g$, which leads to a high density of states near the $\Ef$ of the undoped material. 
These states have been studied in detail by several authors (see \cite{CastroNeto-review} and refs. therein) including the recent proposal of a novel quantum spin Hall effect in the presence of spin-orbit coupling \cite{Kane2005}.
When next-to-nearest neighbors hopping is taken into account, the edge states become dispersive---they acquire a finite $v_g$---and more stable \cite{Sasaki2006}. In addition, electron-electron interactions lead to a magnetic ordering of the edge states \cite{Fujita1996,Son2006} and the appearance of an energy gap in the band structure \cite{Son2006}.
Since the resulting edge states are then spin-polarized, several groups have proposed to use them for spintronics applications such as creating pure spin currents \cite{Wimmer2008} or inducing half-metallic behavior with electric fields \cite{Son2006}.

Here, we analyze electron transport through a ZGNR with a potential well (PW) created by external gates and tuned in such a way that transport inside the well is governed only by the edge states. In this case, while the current flow is essentially homogeneous outside the PW region, it flows along the edges inside it. We show then that the system behaves as a two-paths interferometer, even though the ZGNR is structurally homogeneous, an effect unique to the ZGNR band structure. 
Interference between the two paths can  be tested by either using a magnetic or a transverse electric field to tune the orbital phase difference between the two branches. 

Furthermore, since the ground state corresponds to an antiferromagnetic ordering of the polarization of the two edges, each path corresponds to a different spin orientation. Then, if the initial spin polarization of the incoming electron, set for instance by a ferromagnetic contact, is perpendicular to the intrinsic polarization of the ribbon, the two-paths interference leads to a \textit{rotation} of the spin of the carriers whose angle can be controlled externally, offering an interesting potential for spintronics.

We describe the ZGNR in the tight-binding approximation. The Hamiltonian then reads $H=H_\mathrm{GNR}^0+H_\mathrm{ext}+H_\mathrm{int}$, where
\be
 H_\mathrm{GNR}^0\smeq-t\sum_{\langle i,j\rangle,\sigma}  b^\dagger_{j\sigma} a^{}_{i\sigma}
-t'\!\!\sum_{\langle\langle i,j\rangle\rangle,\sigma}\! \left(a^\dagger_{i\sigma} a^{}_{j\sigma}\smpl b^\dagger_{i\sigma} b^{}_{j\sigma}\right)\smpl \mathrm{h.c.}
\ee
describes the ribbon. Here, $a^\dagger_{i\sigma}$ ($b^\dagger_{i\sigma}$) creates an electron on a Wannier orbital centered at site $\bm{r}_i$ of the sublattice A (B) with spin $\sigma$, $t\simeq2.8$eV and $t'\simeq-0.1t$ \cite{CastroNeto-review} are the nearest- and next-to-nearest- neighbors hopping parameters, respectively. The symbols  $\langle\dots\rangle$ and $\langle\langle\dots\rangle\rangle$ restrict the sum to the corresponding neighboring sites. The borders contain $A$ sites on one edge and $B$ sites on the other. $H_\mathrm{ext}$, which describes the action of external gates, is defined below.
Finally, $H_\mathrm{int}$ describes the electron-electron interaction. 
Because of the high density of states induced by the edge states, the system is magnetically unstable. DFT and Hartree-Fock calculations \cite{Wunsch2008,Son2006} show that the ground state corresponds to an antiferromagnetic ordering of the sublattices' magnetization. Since the latter is mainly localized at the edges, and to capture the essence of this effect, we take into account such interaction by introducing an effective magnetic field only at the edges sites,
\be
H_{\mathrm{int}}=-\mu_BB_a\sum_{\alpha\sigma} \sigma \, a^\dagger_{\alpha\sigma} a^{}_{\alpha\sigma}-\mu_BB_b\sum_{\beta\sigma}\sigma\,b^\dagger_{\beta\sigma}b^{}_{\beta\sigma},
\ee
where $\alpha$ ($\beta$) labels the top (bottom) edge. We take this  field to be perpendicular to the plane of the ZGNR ($\hat{z}$ axis). In the ground state the two edges have opposite magnetizations, $B_b\smeq-B_a$. The value of $B_a$ should, in principle, be determined by a self-consistent calculation. Since its precise value depends on the chemical passivation of the edges \cite{Koskinen2008,Lee2009}, and in order to discuss different situations, we take it here as a free parameter \cite{pristine}.

\begin{figure}[t]
 \includegraphics[width=.3\textwidth,clip]{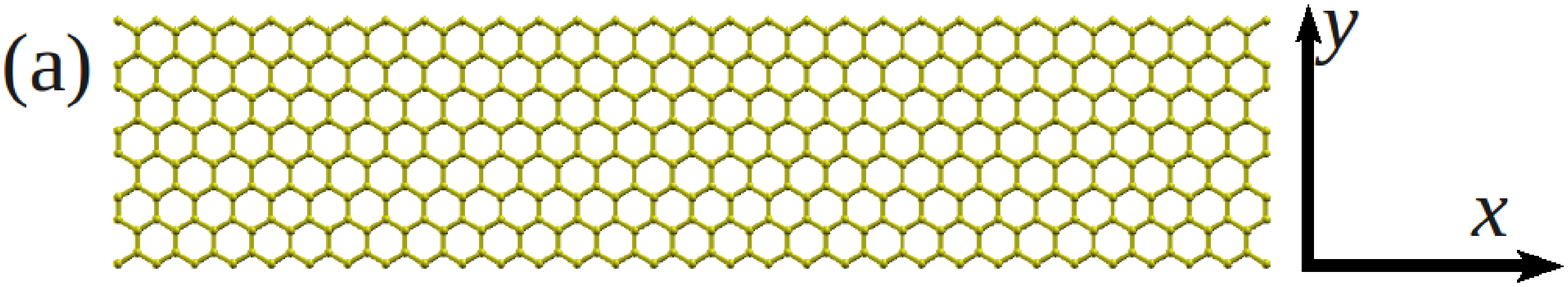}\\
\vspace{.1cm}
 \includegraphics[width=.4\textwidth,clip]{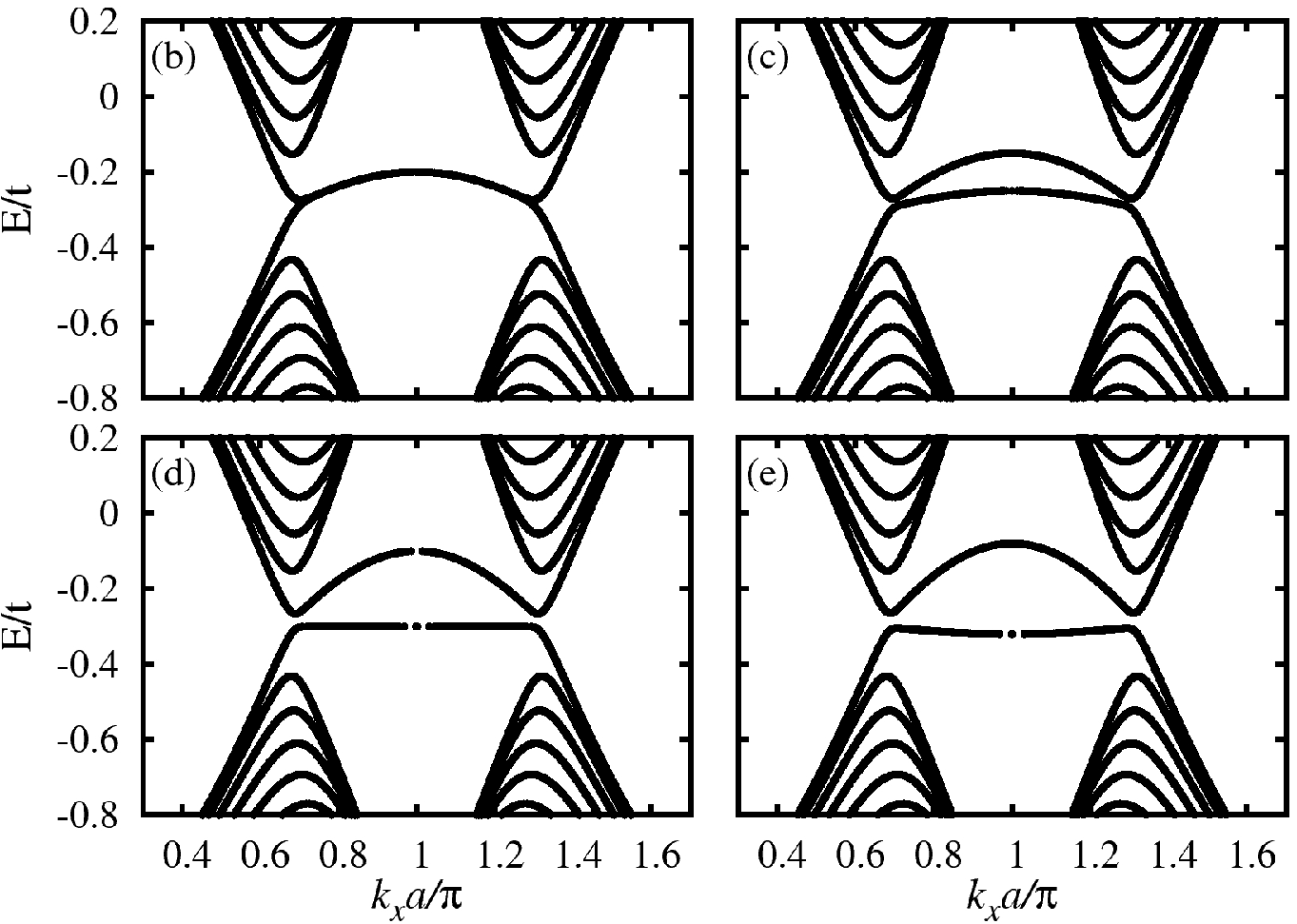}
\caption{(a) Scheme of a ZGNR. The energy of a $32-$ZGNR as a function of the wavevector along the $\hat{x}$ axis is shown for: (b) $B_a=0$; (c) $B_a=t'/2$; (d) $B_a=t'$, (e) $B_a=1.2 t'$. 
 The bands connecting the two non-equivalent Dirac points correspond to the edge states.}
\label{dispersion}
\end{figure}

Figure \ref{dispersion} shows the energy dispersion of a $32$-ZGNR \cite{NZGNR} for different values of $B_a$. Several bands originated from the quantization  along the $\hat{y}$ axis are clearly visible. The bands in the range $k_xa\in[2\pi/3,4\pi/3]$ that are close to the Dirac point, $E\approx3t'$, are the ones that correspond to the edge states with a characteristic localization length $ \lambda(k_x)\simeq -3a_{0}/2\ln|2\cos(k_xa/2)|$\cite{Sasaki2006}. Here, $a=\sqrt{3}a_{0}$ is the lattice parameter with $a_{0}$ the $C$-$C$ bond length. For $B_a=0$ (Fig. \ref{dispersion}b), there are two of those bands (for each spin orientation) that are almost degenerated---there is an exponentially small gap between them. They essentially correspond to the symmetric and antisymmetric combination of the exponentially decaying solutions of each individual edge. For $B_a\neq0$ (Figs. \ref{dispersion}c,\ref{dispersion}d,\ref{dispersion}e), both the spatial and the spin degeneracies are broken. For each spin orientation, each band now corresponds to states localized on a different edge. 
The energy dispersion is approximately given by $E(k_x)\simeq 3t'+(t'\pm\mu_B B_a)(2\cos k_xa+1)$---note that it is nonzero due to the nonzero value of either $t'$ or $B_a$ \cite{Sasaki2006,ferro}.
\textit{The key point is to notice that, for a given energy, the states with opposite spin polarization in the $\hat{z}$ direction are localized on opposite edges of the ZGNR.}   

\begin{figure}[t]
 \includegraphics[width=.4\textwidth,clip]{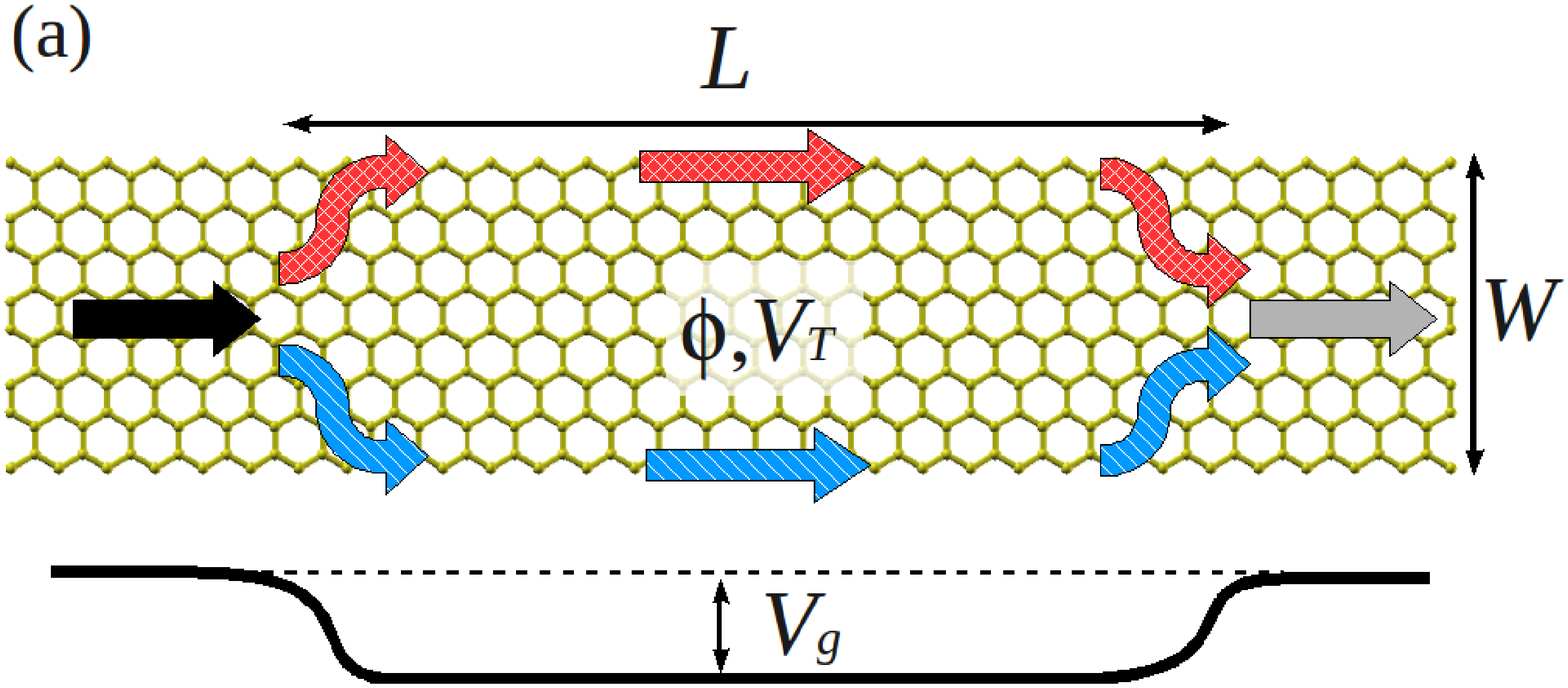}\\
\vspace{.3cm}
 \includegraphics[width=.35\textwidth,clip]{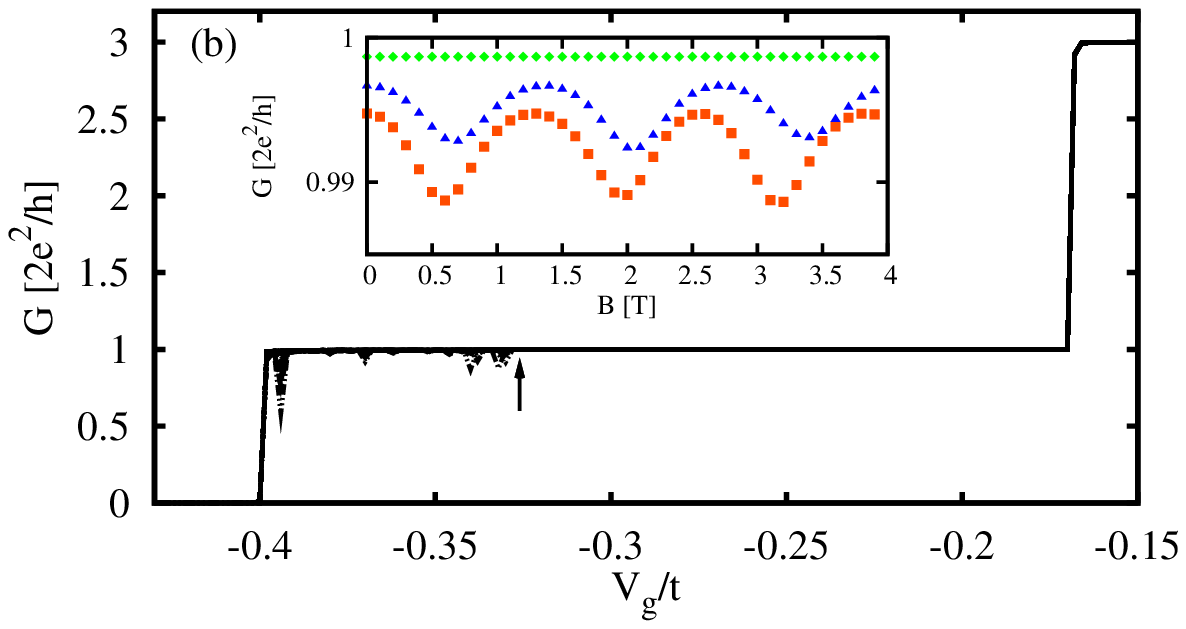}
\caption{(Color online) (a) Schematics of the proposed setup and the potential profile (which depends only on $x$). The two-paths character of the current flow allows for interference effects to manifest; (b) Conductance  of a $32$-ZGNR ($W=45a_0$) as a function of $V_g$ for different values of $B_\perp$ and $\Ef=-0.6t$, $L=2400a$, $\Delta=30a$ and $B_a=B_b=0$. $G$ only changes in the presence of edge states. Inset: $G$ as a function of $B_\perp$ for $V_g/t=-0.384$ ($\blacksquare$), $-0.356$ ($\blacktriangle$) and $-0.31$ ($\blacklozenge$).}
\label{scheme}
\end{figure}

Let us now consider the transport properties of a ZGNR in the presence of an electrostatic potential created by external gates,  
\be
H_\mathrm{ext}=\sum_{i,\sigma} V_g f(x_i) \left(a^\dagger_{i\sigma} a^{}_{i\sigma}+ b^\dagger_{i\sigma}b^{}_{i	\sigma}\right),
\label{Hext}
\ee
where $f(x)$ is a smooth function describing a PW of height $V_g$ (see Fig. \ref{scheme}a). For simplicity, we use a sum of Fermi functions, with the parameter $\Delta$ playing the role of the temperature, to set the spatial profile of $f(x)$ (which depends only on $x$).
We assume that $\Ef<3t'<0$ far from the PW which ensures that the current carrying states in that region are extended throughout the entire width of the ribbon. On the other hand, $V_g<0$ can be tuned in such a way that $\Ef-V_g$ corresponds to the energy of an edge state.  
For the sake of simplicity, we discuss first the conceptually simpler case $B_a=0$ \cite{ferro}.
Then, if $f(x)$ changes smoothly 
, the electrons' wavefunction will adiabatically change from extended to localized, while keeping its band index and having a position dependent wavevector $k_x(x)$. Correspondently, the charge flow will `split' in two paths inside the well and merge again afterwards, creating a `hole' in its spatial distribution (Fig. \ref{scheme}a). In this way, we have create an interferometer which can be tested by introducing a relative phase difference between the two paths.

As the Aharonov-Bohm effect provides the simplest way to do this, we introduce a magnetic field $B_\perp$ perpendicular to the ZGNR (via a Peirls substitution in the hoppings) and calculate the zero temperature conductance using the Landauer approach \cite{Ferrybook}.
For that, we separate the system into a central region (containing the PW) and the lead regions and use the standard recursive method to obtain the lattice Green functions \cite{Ferrybook,Pastawski2001} and the transmission coefficient from them.
Figure \ref{scheme}b shows the conductance $G$ of a $32$-ZGNR as a function of the $V_g$ for different values of $B_\perp$. 
It is apparent that $G$ changes with $B_\perp$ only when $V_g$ is below the threshold where the edge states participate on transport (indicated by the arrow). The inset shows the oscillatory behavior of $G$ as function of $B_\perp$ for three different values of $V_g$. The period is roughly $\phi_0/A'\simeq1.3$T with $\phi_0$ the flux quantum and $A'\simeq L_\mathrm{eff}W$ with $L_\mathrm{eff}\simeq(L-4\times3.5\Delta)$. An increment of the period, due to the reduction of the effective `hole' area, is difficult to see since the visibility of the oscillations is rapidly lost. In addition, and despite this seemly simple picture, the behavior of the conductance is more involved as it shows pronounced narrow dips when $B_\perp\neq0$. This is related to the fact that bonding and antibonding bands are mixed by $B_\perp$ (recall that for $B_a=0$ the gap is exponentially small) and then both bands get involved in transport which in turns leads to Fano-like interference between them \cite{Guzman2009}.

A more interesting situation occurs for $B_a\neq0$. As we mentioned above, in this case, both the spatial and the spin degeneracy are broken. Therefore, an incoming electron with its spin quantize along the $\hat{z}$ axis, will follow either the upper or lower path (colored arrows in fig. \ref{scheme}a) depending on whether its spin is `up' or `down'.
Clearly, in this case there is no interference and the conductance is independent of $B_\perp$. 
Nevertheless, it can be readily verified that if the incoming electron is polarized in the $\hat{n}=\cos\varphi\,\hat{x}+\sin\varphi\, \hat{y}$ direction---its spin state being denoted by $|\hat{n}\up\rangle$---it will be rotated
\be
|\mathrm{in}\rangle=\frac{|\hat{z}\uparrow\rangle+e^{\ci \varphi}|\hat{z}\downarrow\rangle}{\sqrt{2}}\longrightarrow |\mathrm{out}\rangle=\frac{|\hat{z}\uparrow\rangle+e^{\ci (\varphi+\xi)}|\hat{z}\downarrow\rangle}{\sqrt{2}},
\ee
where $\xi$ is the relative phase of the transmission amplitude of the two paths. Due to the symmetry of the setup, the spin projection remains on the plane of the ZGNR. The probability for an electron to keep its spin orientation is  $\cos^2(\xi/2)$ and so we expect  the conductance between two collinear ferromagnetic leads \cite{Tombros2007} to oscillate as a function of $\xi$. Note that we have assumed that $L\ll L_\mathrm{corr}$, where $ L_\mathrm{corr}$ is the spin correlation length of the ferromagnetic order along each edge \cite{Yazyev2008}.
\begin{figure}[t]
 \includegraphics[width=.45\textwidth,clip]{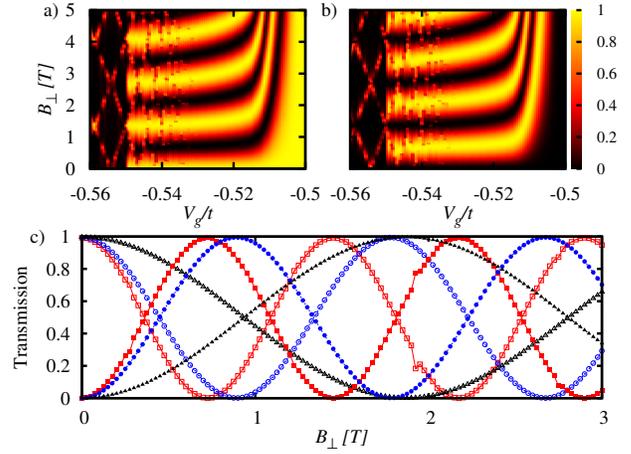}\\
\caption{(Color online) a) Density plot of the spin-resolved transmission $T_{++}$ as a function of the depth of the potential well $V_g$ and the  perpendicular magnetic field $B_\perp$ for a $32$-ZGNR and $\mu_B B_a=t'/2$, $\Ef=-0.8t$, $L=2400 a$, $\Delta=30a$; b) same for $T_{-+}$; c) magnetic field dependence of $T_{++}$ (open symbols) and $T_{-+}$ (filled symbols) for $V_g/t=-0.53$ (squares), $-0.515$ (circles) and $-0.51$ (triangles).}
\label{rotationB}
\end{figure}

Figure \ref{rotationB} shows the spin-resolved transmission probability $T_{\sigma+}$ for an incident electron with spin $|+\rangle=|\hat{x}\up\rangle$ to be transmitted with spin $\sigma=\pm$ (in the same axis) as a function of $V_g$ and $B_\perp$. The relative phase of the two paths is $\xi=2\pi \phi/\phi_0$, where $\phi=B_\perp A_\mathrm{eff}$ is the magnetic flux enclosed by the current flow and $A_\mathrm{eff}=L_\mathrm{eff}W_\mathrm{eff}$ is the effective area. For our geometry, the latter depends mainly on the effective width $W_\mathrm{eff}(V_g)$, which is a function of $V_g$ through the energy dependence of $\lambda(k_x)$ ($W_\mathrm{eff}\simeq W[\coth(W/\lambda)-\lambda/W]$ for $\lambda/W\ll1$).  As expected, the transmission is an simple oscillatory function of $B_\perp$. Note that the shorter period corresponds to the maximum effective area, $\phi_0/(L_\mathrm{eff}W)\simeq1.5$T and that for $V_g>V_g^*\simeq-0.5t$ (threshold for the participation of the edge states) there are no oscillations. The total transmission $T=T_{++}+T_{-+}$ is constant, implying that the effect of the field is to produce a pure spin rotation. It is worth pointing out that $B_\perp$ can not be too large to avoid a transition to a ferromagnetic state.  

\begin{figure}[t]
 \includegraphics[width=.45\textwidth,clip]{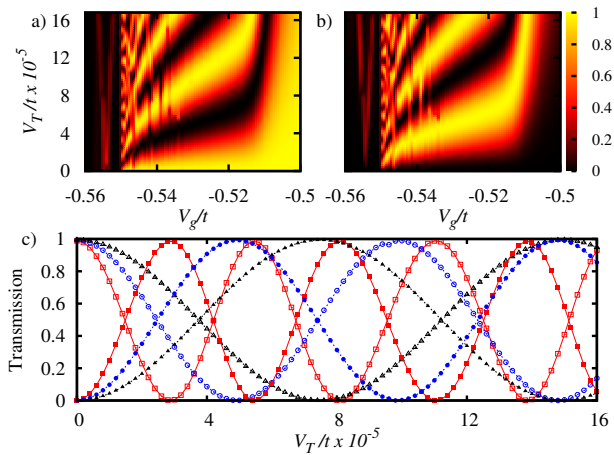}\\
\caption{(Color online) a) and b) Same as Fig. \ref{rotationB} but 
as a function of the transverse electrostatic potential $V_T$.
; c) transverse electric field dependence of $T_{++}$ (open symbols) and $T_{-+}$ (filled symbols) for $V_g/t=-0.544$ ($\square$,$\blacksquare$), $-0.53$ ($\circ$,$\bullet$) and $-0.515$ ($\vartriangle$,$\blacktriangle$).}
\label{rotationVT}
\end{figure}

Interestingly enough, there is also a way to produce a controlled spin rotation using an \textit{all-electrical-setup}. The key is to change $\xi$ by inducing a difference between the wavevectors of the two paths, and therefore changing their relative plane wave phase. This can be achieved by applying a small transverse electrical field that changes the energy of the two paths, and then the wavevectors, in a small fraction and in opposite directions---note that only a change $\delta k_x\simeq 2\pi/L$ is required. The transverse potential is described by 
adding a term $V_T\, ([y_i-W/2]2/W)f(x_i)$ to $V_g f(x_i)$ in Eq. (\ref{Hext}).
Figure \ref{rotationVT} shows the spin-dependent transmission for this setup. As for the previous case, there are clear oscillations indicating the rotation of the spin of the carriers, even for a very small transverse field $E_T=2V_T/W$ ($\simeq2\mu V/$\AA\, for $V_T\simeq 5\, 10^{-5} t$) . Again, the rotation disappears for $V_g\!>\!V_g^*$. The period of the oscillations is in good agreement with the estimated value  $\xi \smeq(k_x^+\smmi k_x^-)L_\mathrm{eff}$ where $k_x^\eta$ is the wavevector of the edge states with energy $\Ef-V_g+\eta\langle V_T\rangle$ and $\langle V_T\rangle$ is the average value of the transverse potential in the corresponding edge state ($\langle V_T\rangle\simeq V_T[\coth(Na/\lambda)-\lambda/Na]$ for  $\lambda/W\ll1$).

Adiabatic transport is not possible for $\mu_B B_a>t'$ as electrons reach a point where $v_g\simeq0$ before they penetrate the well and are then reflected. However, transport is still possible due to a resonant mechanism that involves the upper `w'-shaped edge states band (Fig. \ref{dispersion}). This involves a Landau-Zener like transition between bands, so that the width of the resonances increases as the potential profiles is more abrupt. Some of those resonances are already apparent in Figs. \ref{rotationB} and \ref{rotationVT}. We note that in Fig. \ref{rotationB}, they present a period of $2\phi_0/A_\mathrm{eff}$. This is also present when the incident electron has its spin direction in $\hat{z}$, where we would have naively expected no dependence with $B_\perp$ as the electrons in that case follow a single path. This is, however, not true as each minimum of the `w'-shaped band involves edge states for electrons moving in one direction but  extended states for those moving in the opposite direction---another unique characteristic of the ZGNR band structure. The phase difference is then related to half the area of the PW.
A detailed analysis \cite{Guzman2009} shows that the spin rotation is still possible for some of the resonances, showing that the effect is robust against the precise value of $B_a$.

In summary, we showed that ZGNRs present interesting interference phenomena in the presence of a PW. Moreover, the spin-dependent structure of the edge states allows for a controlled rotation of the spin of the carriers by either magnetic or electric fields. Since the characteristic of the zigzag termination seems to be generic \cite{Akhmerov2008} and robust against disorder \cite{Wimmer2008}, we expect these effects to manifest in less ideal samples, opening a new alternative for spintronics in graphene. 

We thank C. A. Balseiro for useful discussions and P. S. Cornaglia for a careful reading of the manuscript.
We acknowledge financial support from ANPCyT Grant No 483/06 and CONICET PIP 5254/05.


\begin{thebibliography}{31}
\bibitem[{pri()}]{pristine}
\bibinfo{note}{For $B_a=0$, we take the case of pristine graphene as the
  starting point. For hydrogen passivation, the curvature of the energy
  dispersion changes sign so that a potential barrier should be considered in
  that case.}
\bibitem[{NZG()}]{NZGNR}
\bibinfo{note}{Following the standard notation, we identify a given GNR by the
  number $N$ of zigzag chains it contains in the transverse direction, see
  \cite{CastroNeto-review} for details.}
\bibitem[{fer()}]{ferro}
\bibinfo{note}{For ferromagnetic order between edges, $B_a=B_b$, there is a
  spin splitting of the bands but the spatial symmetry is not broked.
  Oscillations similar to the ones discussed below for $B_a=0$ should also
  appear in this case.}
\end{thebibliography}
\end{document}